\documentclass[11pt,a4paper]{article}
\usepackage{jcappub,physics}

\title{Statistics of CMB polarization angles}

\author[a]{James Creswell,}
\author[a,b]{Nadia Dachlythra,}
\author[a,c]{Hao Liu,}
\author[a]{Pavel Naselsky}
\author[a]{and Per Rex Christensen}

\affiliation[a]{Niels Bohr Institute, University of Copenhagen, Blegdamsvej 17, DK-2100 Copenhagen, Denmark}
\affiliation[b]{The Oskar Klein Centre, Department of Physics, Stockholm University, AlbaNova, SE-10691 Stockholm, Sweden}
\affiliation[c]{Key laboratory of Partcle and Astrophysics, Institute of High Energy Physics, CAS, 19B YuQuan Road, Beijing, China}

\emailAdd{james.creswell@nbi.ku.dk}
\emailAdd{konstantina.dachlythra@fysik.su.se}
\emailAdd{liuhao@nbi.dk}
\emailAdd{naselsky@nbi.dk}
\emailAdd{perrex@nbi.dk}

\abstract{We study the distribution functions of the CMB polarization angle $\psi$, focusing on the Planck 2018 CMB maps. We extend the model of Preece \& Battye (2014) of Gaussian correlated $Q$ and $U$ Stokes parameters to allow nonzero means. When the variances of $Q$ and $U$ are equal and their covariance and means are zero, the polarization angle is uniformly distributed. Otherwise the uniform distribution will be modulated by harmonics with $2 \psi$ and $4 \psi$ phases. These modulations are visible in the Planck 2018 CMB maps. Furthermore, the mean value of $U$ is peculiar compared to the power spectrum.}

\keywords{CMBR polarisation, CMBR theory}

\begin{document}

\maketitle

\section{Introduction} \label{sec:intro}

The main task of the fourth generation of CMB experiments (following COBE, WMAP, and Planck) is to detect and constrain cosmological gravitational waves, which are a direct probe of inflation \cite{Ade:2015lrj}. 
The goal of this phase is a reliable measurement of the B mode of polarization, which will require hardware solutions, as well as a modification of the methods for processing and analyzing observational data, including diffuse and point-like foregrounds and effects of systematics.

The polarization angle $\psi$, defined in terms of the Stokes parameters $Q$ and $U$ by 
\begin{equation}
    \psi = \frac{1}{2} \arctan(\frac{U}{Q}),
\end{equation}
has been used as a tool in previous works on CMB polarization.
In~\cite{Vidal:2014ala}, it was shown that measurements of the polarization angle can be used as part of a method for reducing bias in polarization amplitude. Foreground modeling relies on the polarization angle, especially for characterizing the behavior and structure of synchrotron and thermal dust emission \cite{Aghanim:2016cps}.
The possibility of detecting systematics using the polarization angle and its probability density function was recently discussed in~\cite{Preece:2014qaa}.

Unlike the E and B modes of polarization, the Stokes parameters are not rotationaly invariant.
However, it is important to stress, that component separation and extraction of the primordial CMB polarised signal usually takes place in $Q,U$ domain and only then is converted into E/B components.
This conversion is not local and for different masks is accompanied by the presence of different E-to-B leakage terms \cite{Bunn:2002df,Bunn:2010kf,Liu:2018kut,Liu:2019zrt,Liu:2019scs}.

The purpose of this paper is to study the statistical properties of the derived ``cosmological'' polarization signals in  Planck 2018 and, in particular, the statistics of the polarization angle after masking.
Note that Planck CMB maps are contaminated by foreground residuals and systematic effects, and they cannot be described as a pure CMB signal. Therefore any departure of the statistical properties of these maps from theoretical expectations or simulations will be an indication of the presence of such components.
As we will show below, the polarization angle has specific statistical properties when the underlying $Q$ and $U$ data are Gaussian.
Morevover, such properties are independent of the reference frame.
Departures from Gaussianity will be visible in distribution of polarization angles. 
This is why this method provides an important new test in the search of non-Gaussianity.

Recently discussed in \cite{Preece:2014qaa}, due to non-linearity of the polarization angle, the probability density function $\mathcal{P}(\psi)$ is sensitive to the mean values 
\begin{equation}
    \mu_Q=\frac{1}{N_p}\sum_p Q_p, \quad \mu_U=\frac{1}{N_p}\sum_p U_p
\end{equation}
and the variances
\begin{equation}
    \sigma_{QQ}=\frac{1}{N_p}\sum_p(Q_p-\mu_Q)^2, \quad \sigma_{UU}=\frac{1}{N_p}\sum_p(U_p-\mu_U)^2
\end{equation}
and their covariance $\sigma_{QU}$.
Here the index $p$ stands for each pixel outside the mask and $N_p$ is the total number of pixels outside the mask.
For illustration of the properties of $\mathcal{P}(\psi)$, in \cite{Preece:2014qaa} the primary focus was the WMAP7 polarization data, where strong modulations of the distribution function $\mathcal{P}(\psi)$ with respect to uniformity were detected. 

In this paper we re-examine  the  features of the distribution  function $\mathcal{P}(\psi)$ for the CMB polarization angles based on Planck 2018 data release.
In \cite{Preece:2014qaa}, the modulation of $\mathcal{P}(\psi)$ in the WMAP7 data is due to asymmetry of the moments $\sigma_{QQ}/\sigma_{UU}\neq 1$, and $\sigma_{QU}\neq 0$, and has a harmonic shape proportional to $\sin(4\psi)$ and $\cos(4\psi)$.
Here we will demonstrate that for the Planck 2018 data release, the shape of modulations is significantly different from WMAP7, and it includes modes proportional to $\sin(2\psi)$ and $\cos(2\psi)$ in addition to the $\sin(4\psi)$ and $\cos(4\psi)$ modes.
The origin of the strong $2\psi$ modes in the Planck 2018 CMB maps is the mean values of $Q$ and $U$, which are much stronger than in the WMAP data and also peculiar compared to Gaussian simulations based on the best-fit power spectrum.
This discrepancy exists for the first time in Planck 2018 data release, and it is not so apparent in the Planck 2015 data release.

It is tempting to resolve the problem of anamolus means $\mu_Q$ and $\mu_U$ by subtracting them from observational values of $Q$ and $U$.
However, we should not forget that even foreground-cleaned SMICA, Commander, SEVEM and NILC maps are not free from foreground residuals, instrumental noise and effects of systematics.
Thus, subtraction of the means will lead to artificial residuals in the CMB products.  

The outline of this paper is the following.
In section 2 we present the theory of polarization angles and their distributions, first in the zero-mean case, and then in general.
In section 3 we examine the Gaussianity parameters which are linked with the shapes of the polarization angle distribution, namely the variances, covariances, and means from the 2018 Planck CMB maps, comparing them with simulations. 
In section 4 we calculate the polarization angle distribution from the 2018 SMICA map after masking and after downgrading.
In section 5 we study how the polarization angle distribution alone encodes statistical properties of $Q$ and $U$, and limits on how this information can be extracted. 
A summary and conclusion is given in section 6.

\section{Theory of polarization angles and their distributions} \label{sec:2}

In this section the theoretical distribution function for the polarization angle is derived.
This analysis is conducted in the pixel domain and applies to any sample of pixels, whether the whole sky or a subset of the sky remaining after masking.
The basic assumption is that $Q$ and $U$ are correlated Gaussian variables with possibly nonzero means.
We assume that the reduced covariance matrix of the $Q$ and $U$ Stokes parameters is \cite{Tegmark:2001zv}
\begin{equation}
    \vb*{C}_{QU} = \mqty(\sigma_{QQ} & \sigma_{QU} \\ \sigma_{QU} & \sigma_{UU}),
\end{equation}
and we define the vectors 
\begin{equation}
    \vb*{P} = \mqty(Q \\ U), \quad \vb*{\mu} = \mqty(\mu_Q \\ \mu_U),
\end{equation}
where $\mu_Q$ and $\mu_U$ are the means of $Q$ and $U$.
Then the joint probability density of $Q$ and $U$ is given by
\begin{equation} \label{eq:joint-distr}
    \mathcal{P}(Q, U) \propto \exp(-\frac{1}{2} (\vb*{P} - \vb*{\mu})^T \vb*{C}_{QU}^{-1} (\vb*{P} - \vb*{\mu})),
\end{equation}
where ${}^T$ denotes the transpose.
Given the density of $Q$ and $U$, the distribution of the polarization angle 
\begin{equation}
    \psi = \frac{1}{2} \arctan(U, Q)
\end{equation}
can be calculated by integrating eq.~\eqref{eq:joint-distr} over all possible values of the polarization intensity $I=\sqrt{Q^2+U^2}$.

\paragraph{Zero-means case}

The probability distribution of $\psi$ was studied in \cite{Preece:2014qaa} under the assumption that $Q$ and $U$ have zero means but allowing arbitrary covariance.
The resulting distribution function, in the case that $\mu_Q = \mu_U = 0$, is
\begin{equation} \label{eq:P-zero-mean}
    \mathcal{P}(\psi) = \frac{2 \sqrt{\qty| \vb*{C}_{QU} |}}{\pi f_\sigma(\psi)},
\end{equation}
where $\qty| \vb*{C}_{QU} | = \sigma_{QQ} \sigma_{UU} - \sigma_{QU}^2$ is the determinant of the covariance matrix and we have defined the function
\begin{equation} \label{eq:f-sigma}
    f_\sigma(\psi) = \sigma_{QQ} + \sigma_{UU} + (\sigma_{UU} - \sigma_{QQ}) \cos(4 \psi) - 2 \sigma_{QU} \sin(4 \psi).
\end{equation}
If $Q$ and $U$ are uncorrelated and have equal variances, i.e. $\sigma_{QU} = 0$ and $\sigma_{QQ} = \sigma_{UU}$, then the distribution  $\mathcal{P}(\psi)$ is uniform.
Otherwise, the distribution function has sinusoidal modulations with oscillation frequency $4 \psi$.
To illustrate this, we expand to leading order in the quantities $\sigma_{UU}-\sigma_{QQ}$ and $\sigma_{QU}$,
assuming that the variances are almost the same and the covariance is small but comparable with $\sigma_{UU}-\sigma_{QQ}$.
In this approximation eq.~\eqref{eq:P-zero-mean} becomes
\begin{equation} \label{eq:P-zero-mean-approx} 
    \mathcal{P}(\psi) = \frac{1}{\pi}\left(1 + \frac{(1-R) \cos(4\psi) + \kappa \sin(4\psi)}{(1+R)}\right),
\end{equation}
where we have defined the asymmetry parameters
\begin{equation} \label{eq:R-kappa} 
    R = \frac{\sigma_{UU}}{\sigma_{QQ}}, \quad \kappa = \frac{\sigma_{QU}}{\sigma_{QQ}}.
\end{equation}
If $Q$ and $U$ are not correlated and have equal variances, then $R = 1$ and $\kappa = 0$.

\paragraph{Nonzero-means case}

The distribution functions in eqs.~(\ref{eq:P-zero-mean}) and (\ref{eq:P-zero-mean-approx}) assume that the mean values of $Q$ and $U$ are zero.
However, the distribution function $\mathcal{P}(\psi)$ is strongly sensitive to the means of $Q$ and $U$, and since there is no \emph{a priori} reason to expect $Q$ and $U$ to have zero means in observed maps, it is important to consider cases when this assumption does not apply.
For the general case, allowing both nonzero means and arbitrary covariance, the distribution function is~\cite{pmid24046539}
\begin{equation} \label{eq:P-general}
    \mathcal{P}(\psi) = \frac{2 f_{\mu}(\psi)}{\pi f_\sigma(\psi)^{3/2}} \exp(-\frac{(\mu_U \cos(2\psi) - \mu_Q \sin(2\psi))^2}{f_\sigma(\psi)}) \Bigg( \frac{e^{-f_{\mu \sigma}(\psi)^2}}{f_{\mu \sigma}(\psi)} - \sqrt{\pi} \,\, \mathrm{erfc} \qty(f_{\mu \sigma}(\psi)) \Bigg),
\end{equation}
where $\mathrm{erfc}(x) = 1- \mathrm{erf}(x)$ is the complimentary error function, $f_\sigma(\psi)$ is the same as in eq.~\eqref{eq:f-sigma}, and we have defined
\begin{equation} 
    f_{\mu}(\psi) = \qty(\mu_U \sigma_{QU} - \mu_Q \sigma_{UU}) \cos(2\psi) + \qty(\mu_Q \sigma_{QU} - \mu_U \sigma_{QQ}) \sin(2\psi)
\end{equation}
and
\begin{equation}
    f_{\mu \sigma}(\psi) = \frac{f_{\mu}(\psi)}{\sqrt{\qty|\vb*{C}_{QU}| f_\sigma(\psi)}}.
\end{equation}
When $\mu_Q$ and $\mu_U$ to $0$ are set to zero in the general formula (eq.~\eqref{eq:P-general}) and the indeterminate terms are handled correctly, the result of \cite{Preece:2014qaa}, i.e.~eq.~\eqref{eq:P-zero-mean}, is recovered.
Another useful simplification can be obtained by neglecting $\sigma_{QU}$ and assuming that $\sigma_{QQ} = \sigma_{UU}$, but allowing nonzero means.
In this case, to leading order in $\mu_Q$ and $\mu_U$, we find
\begin{equation} \label{eq:P-zero-cov}
    \mathcal{P}(\psi) = \frac{1}{\pi} + \frac{\mu_Q \cos(2 \psi) + \mu_U \sin(2\psi)}{\sqrt{2\pi\sigma_{QQ}}}.
\end{equation}
The $\sin(2\psi),\cos(2\psi)$ modulation has an amplitude proportional to $\mu_Q$ and $\mu_U$, and a phase related to the relative amplitudes of $\mu_Q$ and $\mu_U$.

Thus, for random Gaussian $Q$ and $U$ with nonzero mean values $\mu_Q,\mu_Q$ and small but nonzero
$|R-1|\ll 1$ and $|\kappa| \ll 1$, we may expect to get two major type
of modulations, proportional to $\sin(2\psi), \cos(2\psi)$ due to
nonzero means, and $\cos(4\psi),\sin(4\psi)$ due to nonzero $R-1$ and $\kappa$. Any departures from these modulations (or significant enhancement of the corresponding amplitudes) will indicate contamination of the derived CMB products from the observational data. In the next sections we will confront the Planck 2018 observational data with these theoretical predictions.

\section{Statistics of Gaussianity parameters in the 2018 Planck maps}

\subsection{Variances and asymmetry parameters}

\begin{figure}[tbp]
    \centering
    \includegraphics{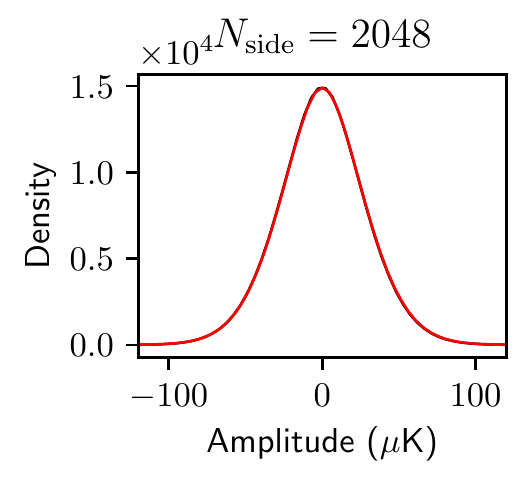}
    \includegraphics{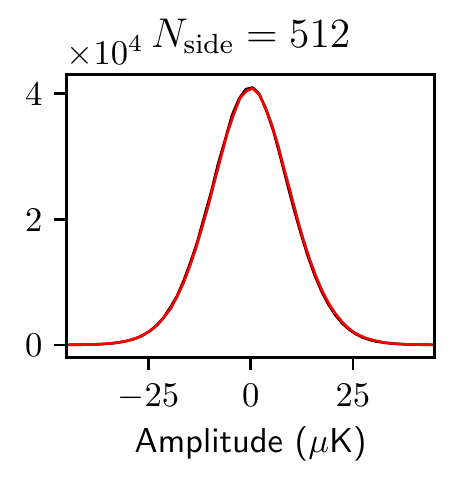}
    \includegraphics{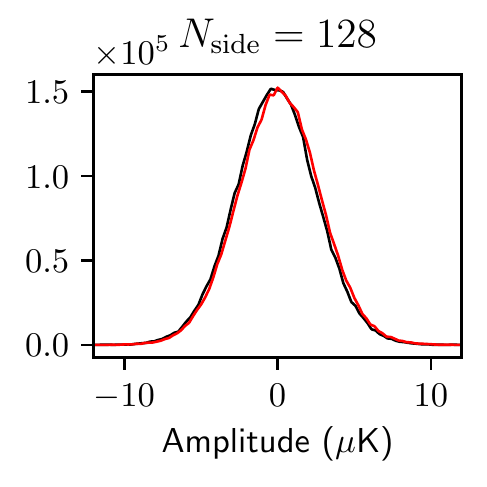}
    \includegraphics{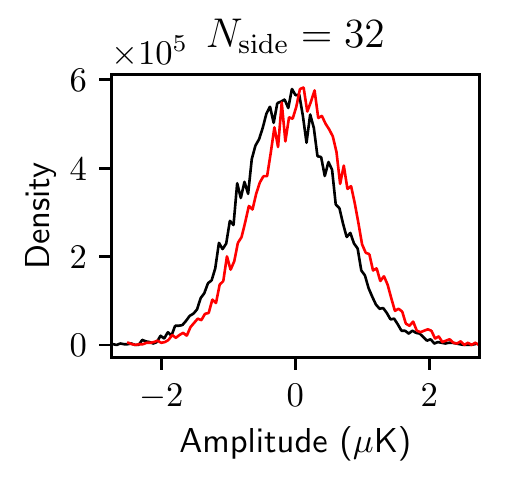}
    \includegraphics{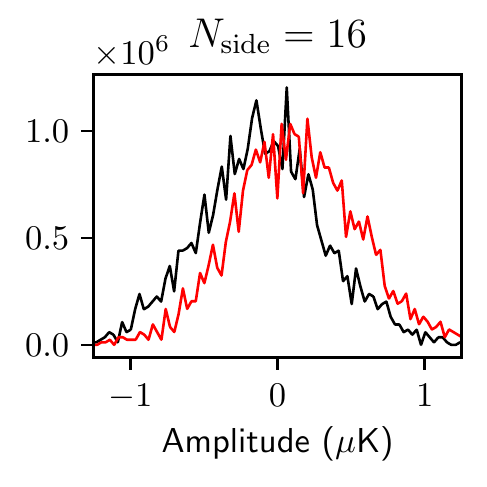}
    \includegraphics{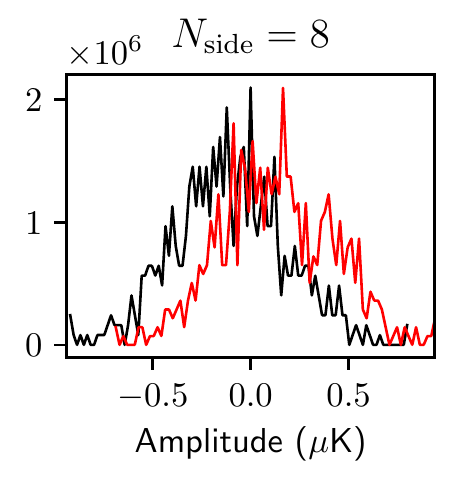}
    \caption{\label{fig:QU-dists} Number of counts versus amplitudes $Q$ (black) and $U$ (red) for 
    Planck 2018 SMICA map with Galactic mask applied for $N_\mathrm{side}=2048,512,128$ (top row) and $N_\mathrm{side}=32,16,8$ (bottom row).}
\end{figure}

As pointed out in the previous section, the shape of the polarization angle distribution function critically depends on the parameters $R$ and $\kappa$ characterizing asymmetry and correlation of $Q$ and $U$.
Here we start with a preliminary analysis of the Planck 2018 products, beginning with the distributions of $Q$ and $U$ for the 2018 SMICA map at different angular resolutions, shown in figure~\ref{fig:QU-dists} \cite{Akrami:2018mcd}.
The distributions are all visibly marginally close to Gaussian, but the means are nonzero and the widths vary with the angular resolution.
Downgrading reduces the variances of $Q$ and $U$ but keeps the means fixed.
Therefore, with decreasing resolution, the nonzero $Q$ and $U$ means become more easily visible and the $Q$ and $U$ distributions separate.

For the SMICA, NILC, Commander and SEVEM maps outside their mask the parameters $\sigma_{QQ}$, $\sigma_{QU}$, $R$  and $\kappa$ are presented in Table \ref{par} for $N_\mathrm{side}=2048$.
From Table \ref{par} one can see that all the CMB products are characterised by $|R-1|\simeq |\kappa|$.
Therefore, we expect their polarization angle distributions to have $\sin(4\psi)$ and $\cos(4\psi)$ modes with comparable amplitudes.

\begin{table}[tbp]
    \centering
        \begin{tabular}{|ccccc|}
            \hline
            Map & $\sigma_{QQ}$ ($\mu \mathrm{K}^2$) & $\sigma_{QU}$ ($\mu \mathrm{K}^2$) & $R$ & $\kappa$  \\
            \hline
            SMICA & $813.552$ & $-3.683$ & $1.0089$ & $-0.0045$ \\
            Commander & $802.189$ & $-2.628$ & $1.0089$ & $-0.0033$ \\
            NILC & $646.368$ & $-1.933$ & $1.0072$ & $-0.0029$ \\
            SEVEM & $436.429$ & $-1.201$ & $1.0076$ & $-0.0028$ \\
            \hline
        \end{tabular}
    \caption{\label{par} $\sigma_{QQ}$, $\sigma_{QU}$, $R$, and $\kappa$ for Planck 2018 CMB products. For all the maps we evaluate the quantities at $N_\mathrm{side}=2048$. The polarization mask is used for all calculations.}
\end{table}

\subsection{Means and comparison with simulations} 

\begin{figure}[tbp]
    \centering
    \includegraphics{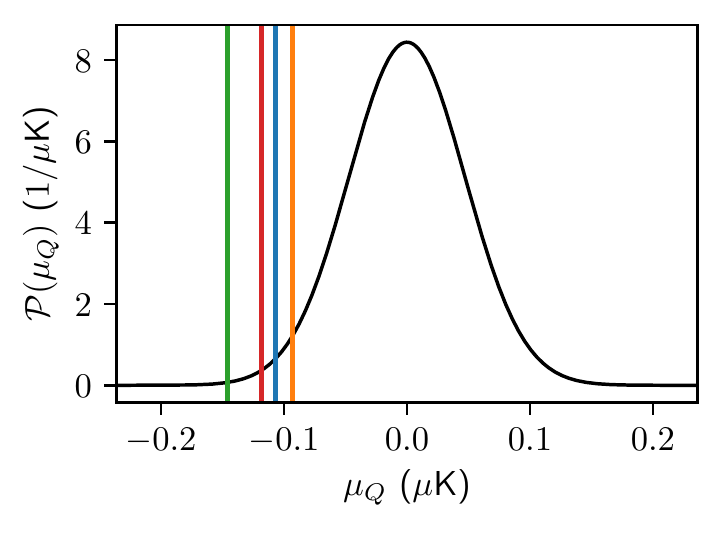}
    \includegraphics{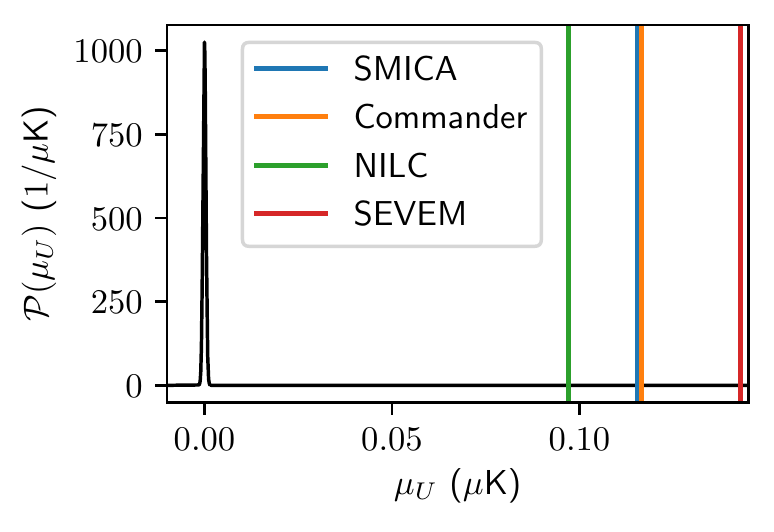}
    \caption{\label{fig:mu-dists} Visualization of the values of $\mu_Q$ (left panel) and $\mu_U$ (right panel) for the four Planck 2018 release CMB maps. In black is the distribution function predicted by the best-fit LCDM power spectrum.}
\end{figure}

For the non-masked sky, the means $\mu_Q$ and $\mu_U$ are nothing but the 
corresponding monopoles, which are usually subtracted from the CMB products. However, for masked skies these means absorb the contribution from low multipoles (quadrupole, octupole etc.) and are
no longer associated with pure monopoles. 
As shown in section \ref{sec:2}, the $\mu_{Q}$ and $\mu_{U}$ means generate $\sin(2\psi),\cos(2\psi)$ modulations of the density function of the polarization angle. For the masked Planck 2018 CMB maps, 
we show in figure~\ref{fig:mu-dists} their values, compared with the probability distribution function for these means predicted by the best-fit LCDM power spectrum \cite{Aghanim:2019ame}.

\begin{table}[tbp]
    \centering
        \begin{tabular}{|ccccc|}
            \hline
            Map & $\mu_Q$ ($\mu \mathrm{K}$) & $\mu_U$ ($\mu \mathrm{K}$) & $p_{Q}$ & $p_{U}$  \\
            \hline
            SMICA
            & $-0.107$ & $0.116$ & $0.024$ & $0.0$ \\ 
            Commander
            & $-0.093$ & $0.116$ & $0.049$ & $0.0$\\ 
            NILC
            & $-0.146$ & $0.097$ & $0.002$ & $0.0$ \\ 
            SEVEM
            & $-0.118$ & $0.143$ & $0.012$ & $0.0$\\ 
            \hline
        \end{tabular}
    \caption{$Q$ and $U$ monopoles taken from the Planck CMB maps and their $p$-values with respect to the distributions for $\mu_Q$ and $\mu_U$ implied by the theoretical LCDM best-fit power spectra. The $p$-values are two-sided, i.e. they give the probability of at least such a deviation from $0$ in either direction.}
    \label{tab:monopoles}
\end{table}

The value of $\mu_Q$ and $\mu_U$ from each map is shown in table~\ref{tab:monopoles}.
We also calculate the $p$-value of each monopole assuming they should be normally distributed with zero mean and standard deviation determined by the LCDM best-fit power spectra (for further discussion of this point, see the conclusion).

The $Q$-monopoles of SMICA, NILC, and SEVEM are in approximate agreement with each other, and are inconsistent with the LCDM $E$-mode spectrum at a significance of $\approx \! 1$--$3\%$.
The $U$-monopoles of all four maps are broadly consistent with each other.
However, they disagree with the LCDM random simulations at a significance of $> \! 250\sigma$.

\begin{figure}
    \centering
    \includegraphics{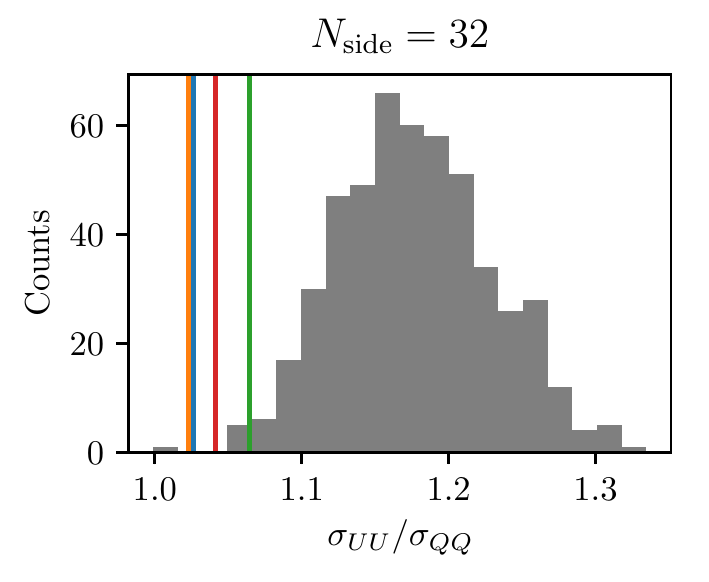}
    \includegraphics{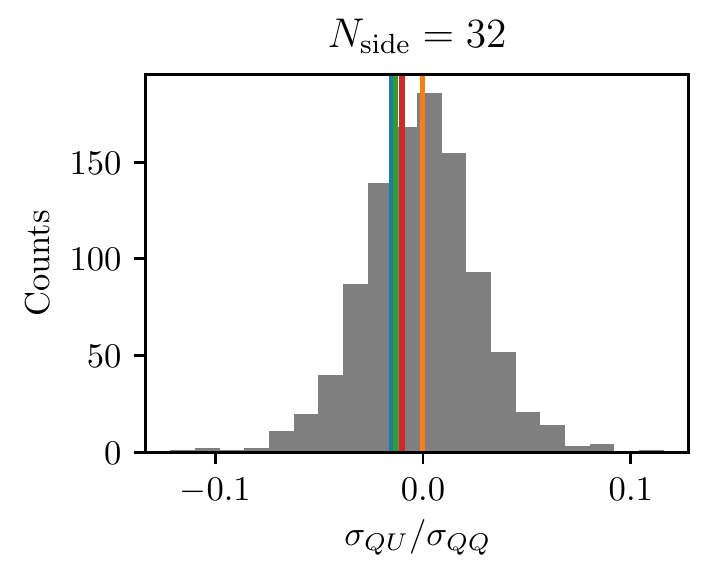}
    \includegraphics{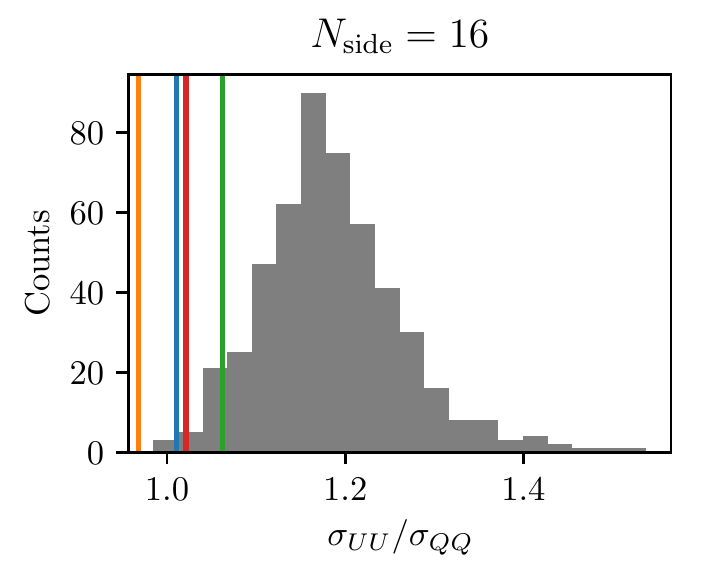}
    \includegraphics{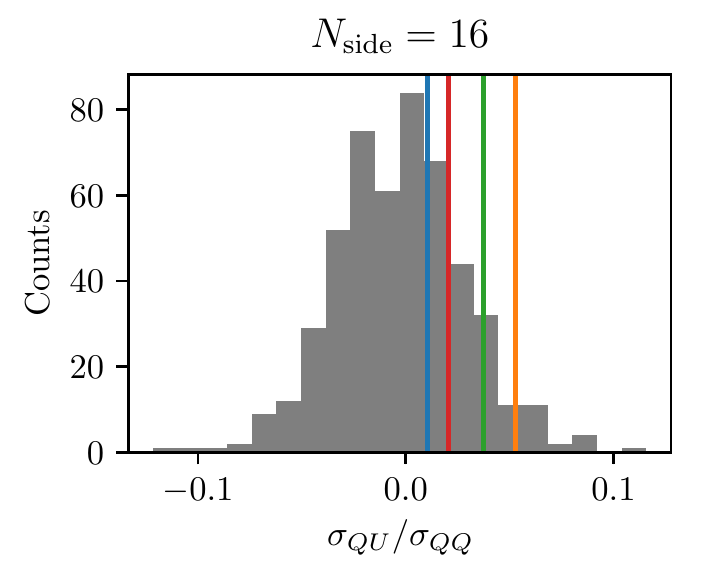}
    \includegraphics{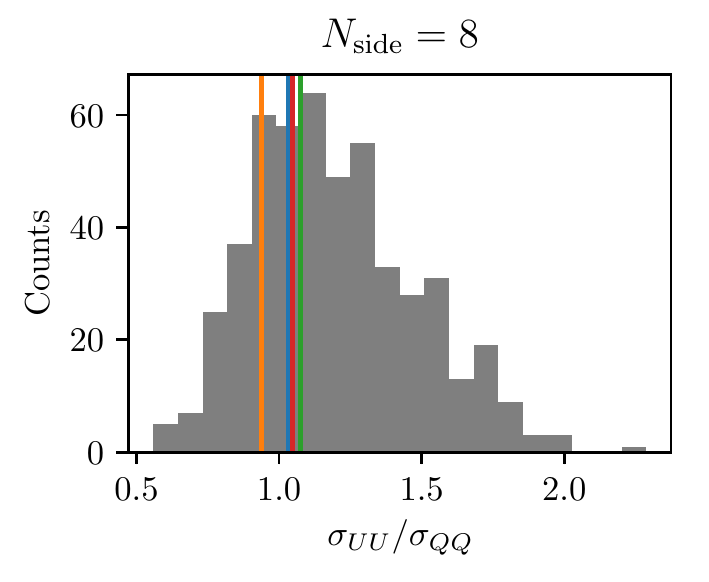}
    \includegraphics{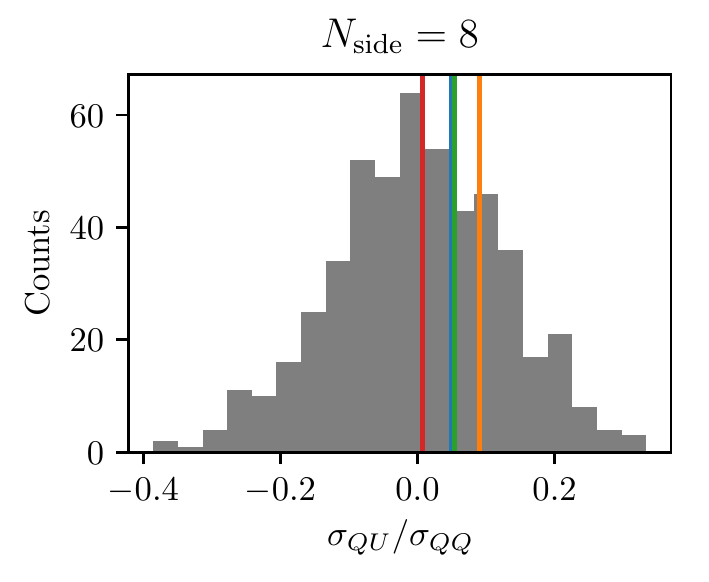}
    \caption{Histograms show values of $R$ and $\kappa$ from 500 FFP9 simulations constructing from the scalar, tensor, and non-Gaussian components with $r = 0.05$ and $f_{NL} = 7$. Vertical bars show the values from SMICA (blue), Commander (orange), SEVEM (green), and NILC (red). We calculate the distributions and values at $N_\mathrm{side} = 32, 16, $ and $8$ (top to bottom). }
    \label{fig:ffp9-distributions}
\end{figure}

For low resolution maps with $N_\mathrm{side}$ from $32$ to $8$ in figure~\ref{fig:ffp9-distributions}, we show the values of $R$ and $\kappa$ taken from 500 FFP9 simulations~\cite{Ade:2015via}.
For $N_\mathrm{side}=16$ and $32$, the parameter of asymmetry for SEVEM map marginally agrees with the FFP9
simulations, while all others reveal obvious peculiarity 
of the CMB maps with respect to FFP9. Needless to mention that FFP9 simulations are biased with preferred value $R\simeq 1.16-1.2$. For random Gaussian simulations all the maps
agree well with $R\simeq 1$, and
the parameter $\kappa$ is in agreement with simulations for all the maps.

\section{Detectability of harmonic modulations in the 2018 Planck maps}

\subsection{Histograms and distribution functions}

For each of the Planck 2018 CMB maps, the distribution function of the polarization angle, $\mathcal{P}(\psi)$, is estimated from the data using a histogram--the number of counts as a function of the binned polarization angle.
The measured histogram, which we denote $\Pi(\psi)$, is then compared to the the theoretical $\mathcal{P}(\psi)$. 
The procedure for constructing the histogram is as follows:
the range from $-\pi/2$ to $\pi/2$ is divided into $k$ equally-sized subintervals.
Then, each pixel on the sky is assigned to the corresponding subinterval in which its value of $\psi$ lies.
Finally, the counts of all bins are normalized by the total number of pixels to yield an estimate of the distribution function.

Practical implementation of this method is the following. We will start from the polarization angle map with given angular resolution (characterized by $N_\mathrm{side}$) with total number of pixels $N_\mathrm{pix}$.
If the theoretical distribution function $\mathcal{P}(\psi)$ (eq.~\ref{eq:P-general}) accurately describes the data in average over a statistical ensemble of realisations, then each realisation of the angular distribution function can be represented as
\begin{equation} \label{chance}
    \Pi(\psi)=\mathcal{P}(\psi)+n(\psi)
\end{equation}
where $n(\psi)$ is a random noise component, perturbing $\Pi(\psi)$ around the theoretical expectation $\mathcal{P}(\psi)$.
Then the whole range of the polarization angle is divided into $k$ bins with
width $\pi/k$. To the $i$-th bin ($i = 1, \dots, k$) are assigned the pixels whose angles $\psi_{i}$  lie in the range
\begin{equation} \label{bin}
    -\frac{\pi}{2} + \frac{(i-1) \pi}{k} \leq \psi_i \leq -\frac{\pi}{2} + \frac{i  \pi}{k},
\end{equation}

Let's denote the average of the distribution function $\Pi(\psi)$ over the bin centered at $\Psi_i$ as
\begin{align}
\left< \Pi(\Psi_i) \right> =
    \frac{k}{\pi}
    \int_{\Psi_i-\frac{\pi}{2k}}^{\Psi_i+\frac{\pi}{2k}} 
     \Pi(\psi) \, d\psi =
 \mathcal{P}(\Psi_i)+  \left< n(\Psi_i) \right>
   \label{bin1} 
\end{align}
where  we have made the approximation that $\mathcal{P}(\psi)$ is constant over the range of the bin, which is valid to the extent that the bins are small and $k$ is large enough. In this case the expectation of
$\left< n(\Psi_i) \right>$ vanishes after average over all bins, while for each bin, the
departure from zero value can be estimated as follows:
\begin{eqnarray}
\left< n(\Psi_i) \right>\sim \sqrt{\frac{k}{N_\mathrm{pix}}}\varepsilon(\Psi_i)
\label{bin2}
\end{eqnarray}
where $\varepsilon(\Psi_i)$ is a random variable, uniformly distributed
within the interval $-\pi/2,\pi/2$ at $k=N_\mathrm{pix}$.
Thus, when $k/N_\mathrm{pix}\ll 1$, the  noise component  for binned distribution will be subdominant in respect to the second  and third terms in eq.~(\ref{bin}) for 
\begin{eqnarray}
k\ll N_\mathrm{pix}(R-1)^2,\hspace{0.2cm} k\ll N_\mathrm{pix}\frac{\mu^2_Q}{\sigma_{QQ}},\hspace{0.2cm} k\ll N_\mathrm{pix}\frac{\mu^2_U}{\sigma_{QQ}}
\label{bin3}
\end{eqnarray}
For $N_\mathrm{pix} \approx 5 \times 10^7$ (i.e. $N_\mathrm{side} = 2048$), we have verified that $k=100$ is a reasonable choice of binning that provides enough bins to have sufficient resolution to see the trend, but also enough smoothing to prevent random fluctuations from destroying the trend.

\subsection{Distributions and effect of smoothing}

\begin{figure}[tbp]
    \centering
    \includegraphics{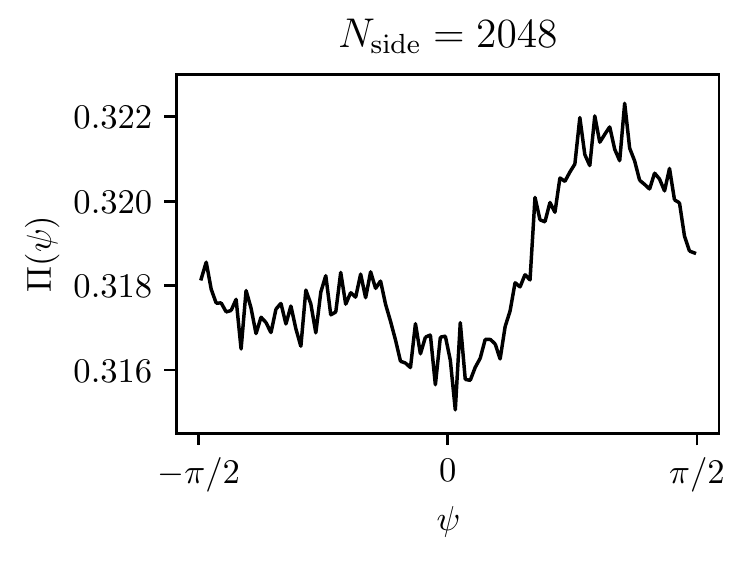}
    \includegraphics{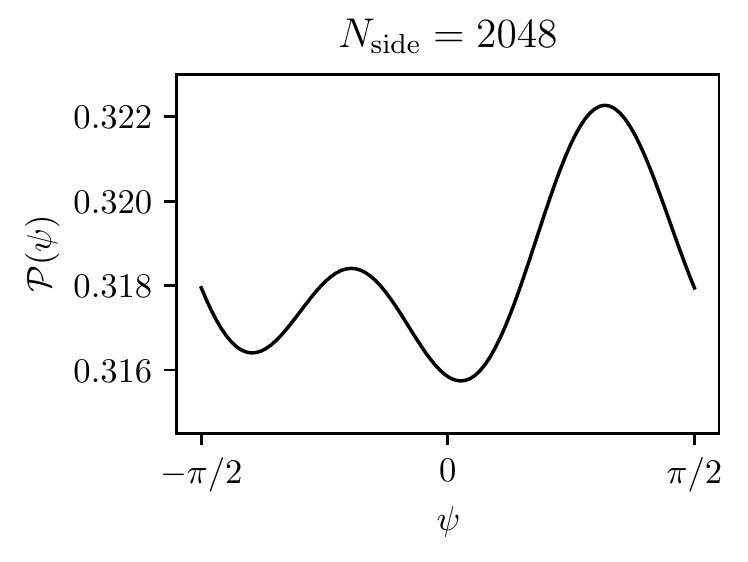}
    \includegraphics{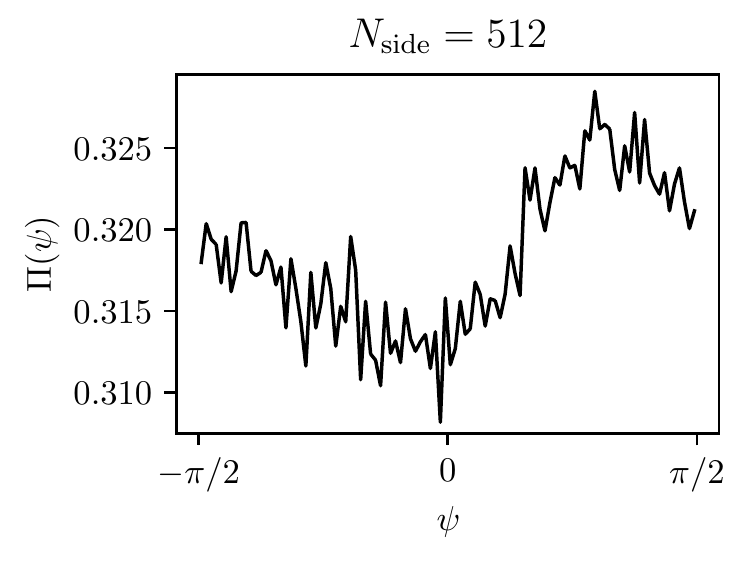}
    \includegraphics{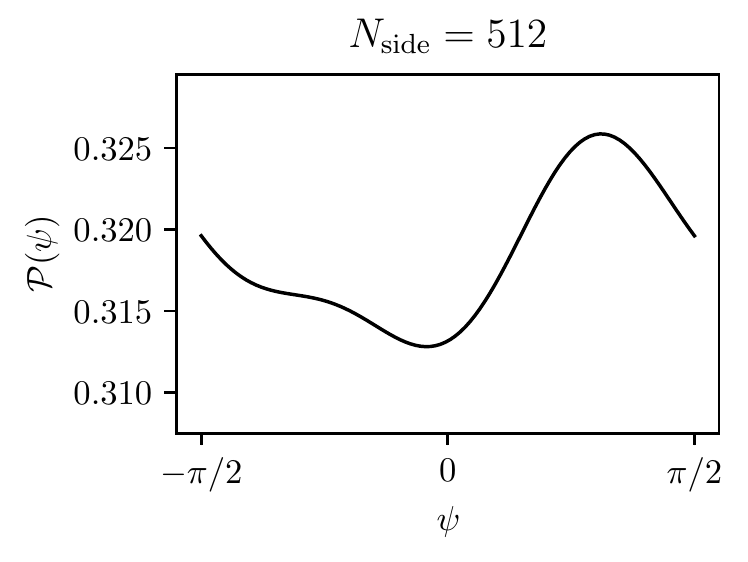}
    \caption{The actual distribution for SMICA outside the mask (left) compared with the theoretical Gaussian distribution using means and variances extracted from the masked data (right). Note enhancement of $2\psi$-mode associated with downgrading from $N_\mathrm{side} = 2048$ (top row) to $N_\mathrm{side} = 512$ (bottom row).}
    \label{fig:dists-actual-theory}
\end{figure}

Followng this procedure, in figure~\ref{fig:dists-actual-theory} we show the binned angular distribution function for the 2018 SMICA map with $N_\mathrm{side}=2048$ and $k=100$ bins.
The recovered histograms agree with the theoretical expectations. 
In the same figure we show the same distribution for $N_\mathrm{side}=512$, again with the same number of bins $k=100$. One can clearly see the enhancement of the $2\psi$ modulations with respect to $4\psi$ modes after downgrading, which is expected.

\begin{figure}[tbp]
    \centering
    \includegraphics{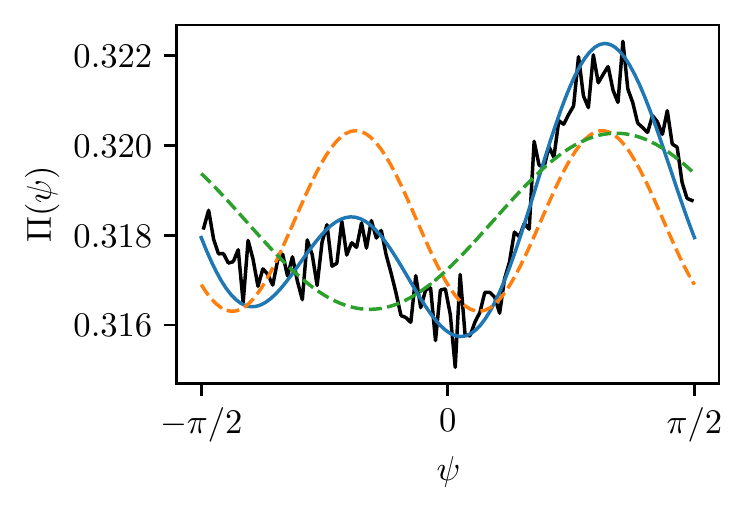}
    \includegraphics{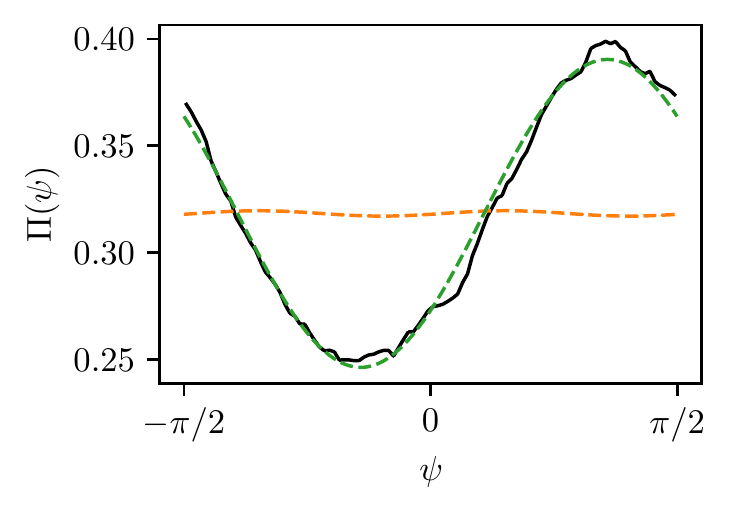}
    \caption{Left panel: distribution function from masked SMICA (black) compared to the theoretical Gaussian model having the same covariance matrix and means (blue), as well as approximate models obtained by neglecting the means (orange, dashed) or the covariance (green, dashed). The distribution function from SMICA is calculated using binning with $k=100$. Right panel: same as left,  but after $1^\circ$ smoothing of the maps. The arbitrary Gaussian model is indistinguishable from the zero-covariance model, and both closely agree with the actual data. }
    \label{fig:dists-smoothing}
\end{figure}

In many aspects of the Planck data analysis, the high resolution $Q$ and $U$ maps (for instance, $N_\mathrm{side}=2048$) are the subjects of filtering by Gaussian kernels with a characteristic scale of smoothing $\Theta$, while the number of pixels
$N_\mathrm{pix}$ is still the same as for unfiltered maps.
When the $Q$ and $U$ maps are smoothed, all the elements of $\sigma_{QU}$ are reduced, while $\mu_Q$ and $\mu_U$ are left unchanged.
Consequently, smoothing will reduce the amplitude of the $4\psi$ modulation relative to the $2\psi$ modulation.
Even a modest amount of smoothing is enough to dramatically alter the shape of $\mathcal{P}(\psi)$.
As an example, we smooth the SMICA maps with a FWHM of $1^\circ$ and recalculate the actual distribution, as well as the theoretical distributions using the smoothed $\sigma_{QU}$, shown in  figure~\ref{fig:dists-smoothing}.
After the $1^\circ$ smoothing, $\mathcal{P}(\psi)$ of SMICA is totally dominated by the $2\psi$-modulation.
This can be understood in the following way:
the amplitude of the $4\psi$ modulation is related to the parameters $R$ and $\kappa$, which are stable when one reduces the resolution of the maps.
Therefore the $4\psi$ terms are of roughly equal amplitude in the unsmoothed and smoothed data.
However, the amplitude of the $2\psi$ modulation is proportional to the mean ($\mu_Q$) divided by the standard deviation ($\sigma_{QQ})$.
When the map is smoothed or downgraded, the mean ($\mu_Q$) is totally unchanged, but the standard deviation is reduced considerably.
Therefore, smoothing or downgrading increases the amplitude of the $2\psi$ terms, while leaving the amplitude of the $4\psi$ terms fixed. Correspondingly, the total distribution function transitions to being dominated by the $2\psi$ modulation, as in
figure~\ref{fig:dists-smoothing}.

The distributions for the other 2018 Commander, NILC, and SEVEM look similar to those of SMICA as shown in figures~\ref{fig:dists-actual-theory} and figures~\ref{fig:dists-smoothing}.
Confer Tables~\ref{par} and \ref{tab:monopoles}, which show that all maps have comparable values of the means and the asymmetry parameters $R$ and $\kappa$.

\section{Extracting Q and U statistics from polarization angle distributions}

The leading order approximations presented in section~\ref{sec:2}, i.e.\ eqs.~(\ref{eq:P-zero-mean-approx}) and (\ref{eq:P-zero-cov}), give rise to a simple procedure in which it is possible to estimate the asymmetry parameters $R$ and $\kappa$, as well as the means divided by the standard deviations of $Q$ and $U$, by fitting $\sin(4\psi), \cos(4\psi), \sin(2\psi), \cos(2\psi)$ curves to the polarization angle distribution.
This can be used as a consistency check and also illustrates the way in which statistical properties of $Q$ and $U$ are encoded into the polarization angle distribution. 
In circumstances when only the polarization angle but not the $Q$ and $U$ data are available, this method could also be used to obtain information about $Q$ and $U$.

\begin{figure}[tbp]
    \centering
    \includegraphics[width=0.45\textwidth]{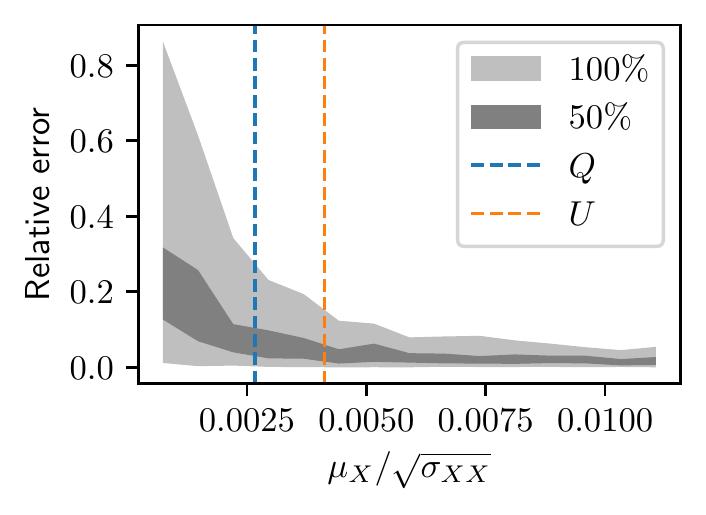}
    \includegraphics[width=0.47\textwidth]{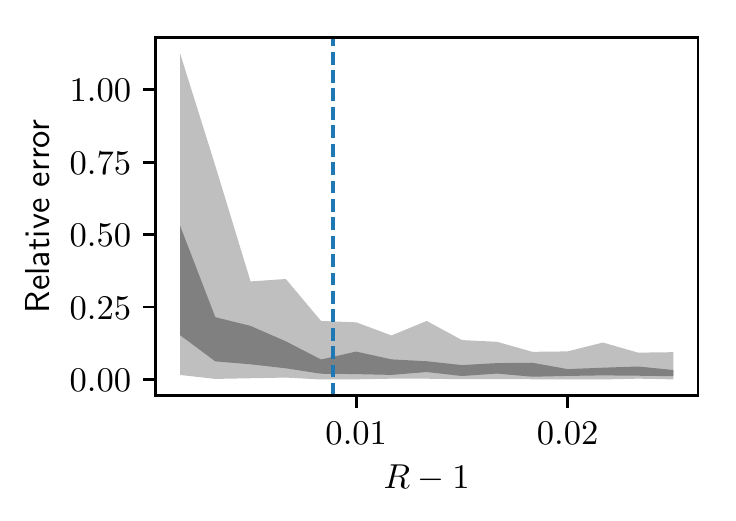}
    \includegraphics[width=0.45\textwidth]{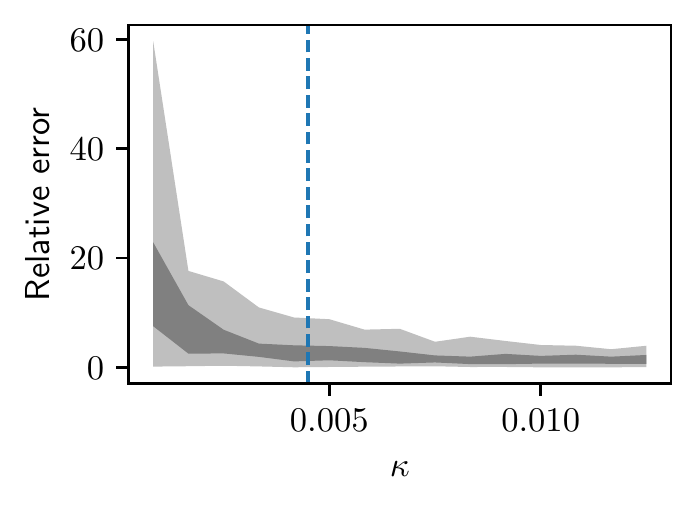}
    \caption{Distribution of relative errors in extraction of $\mu_X/\sqrt{\sigma_{XX}}$ ($X = Q$ or $X = U$), $R$, and $\kappa$ from  the polarization angle distribution $\Pi(\psi)$ by fitting  $\sin(4\psi), \cos(4\psi), \sin(2\psi)$, and  $\cos(2\psi)$ curves to the distribution, as functions of the true input values. The light grey band show the full range of variation of the relative errors from 100 Gaussian simulations; the dark grey band shows the 25th--75th percentile middle range. The colored bars show the corresponding position of the 2018 SMICA map.}
    \label{fig:recovery}
\end{figure}

In figure~\ref{fig:recovery}, we show the results of simulations illustrating the performance of this procedure.
Random Gaussian data are generated with different input means and covariance matrices.
Then we determine the coefficients of the sinusoidal terms using linear least squares regression and compute the relative error of the result with the known input.

Evidently, for maps with properties comparable to SMICA, $\mu$ can be determined from the polarization angle distribution alone with a relative error of approximately $10\%$. $R - 1$ can be extracted with a similar error.
On the other hand, $\kappa$, representing the $Q$--$U$ correlation, is only weakly influential on the polarization angle distribution for maps like SMICA, and the value of $\kappa$ cannot be extracted with any meaningful accuracy in this case.

\section{Discussion and conclusion}

We have investigated the distribution of the polarization angle.
Assuming a correlated Gaussian model for the Stokes parameters $Q$ and $U$, the polarization angles are expected to have a non-uniform distribution.
To leading order, the non-uniformity can be characterized by a superposition of oscillations with frequencies $2\psi$ and $4\psi$, the former arising from nonzero means of $Q$ and $U$, and the latter from unequal variances or nonzero covariance.
The polarization angle is therefore proved a useful diagnostic for these features.

In the work of Preece and Battye \cite{Preece:2014qaa}, histograms of the polarization angle were used to make a simple $\chi^2$ test for systematics in the WMAP 7 data.
We have found similar qualitative results for the WMAP 9 polarization angle distributions, and we have also made plots of the Planck 2018 angle distributions.

The most striking feature found in the Planck 2018 angle distributions is a strong $2\psi$ mode, whose amplitude is expected to be determined by the value of $\mu_U$, which in turn is determined by the $B$-mode power spectrum.
Note that:
\begin{align} 
    \mathrm{var}\qty(\mu_{Q}) &= \sum_{\ell} C_\ell^{EE} \qty(\frac{2 \ell + 1}{\pi})\frac{(\ell - 2)!}{(\ell + 2)!}; \label{eq:varQ} \\
    \mathrm{var}\qty(\mu_{U}) &= \sum_{\ell} C_\ell^{BB} \qty(\frac{2 \ell + 1}{\pi})\frac{(\ell - 2)!}{(\ell + 2)!}. \label{eq:varU}
\end{align}
Therefore the dominant contribution to the variance $\mathrm{var}\qty(\mu_{U})$ is the low-$\ell$ terms of the power spectra $C_\ell^{BB}$.
However, in the actual Planck 2018 data, there is a strong discrepancy between the values of $\mu_U$ in the 2018 Planck maps and the low-$\ell$ terms of the $B$-mode power spectrum.
We note that this discrepancy exists for the first time in the 2018 data release. 
In the 2015 CMB maps, the $Q$ and $U$ monopoles are much smaller and appear that they may have been set to effectively zero by hand, e.g. SMICA~\cite{Adam:2015tpy} has $\mu_Q = 4 \times 10^{-4} \, \mu \mathrm{K}$ and $\mu_{U} = 7 \times 10^{-5} \, \mu \mathrm{K}$.
Moreover, we find that $\qty|\mu_Q| \sim \qty|\mu_U|$ for all 2015 Planck CMB products, whereas the power spectra predict that these quantities should differ by an order of magnitude or so. 

Another important message found in the analysis above is that FFP9 simulations have a bias in terms of the asymmetry parameter $R$ at the level 1 over 500 realisations in respect to correlated Gaussian simulations of $Q$ and $U$ Stokes parameters. This effect requires future investigation of
the properties of FFP9 simulations, since they are widely in use for
investigation of statistical properties of the derived CMB products.

In general, we would like to conclude that even small deviation of the
polarization angle distribution function from uniformity (at the level of 2-10\%) contains very valuable information about possible contamination of the CMB signals. Our method can be successfully implemented for small patches of the sky, like the BICEP2 zone, etc.,
providing new inside on detectability of the primordial B-mode of
polarization from cosmological gravitational waves. In future work, we will extend the theory of polarization angle distributions to the $\psi_E$ and $\psi_B$ angles associated with the $E$- and $B$-modes.

\acknowledgments

Some results of this paper are based on observations obtained with Planck,\footnote{http://www.esa.int/Planck} an ESA science mission with instruments and contributions directly funded by ESA Member States, NASA, and Canada.
Some of the results in this paper have been derived using the HEALPix package~\cite{Gorski_2005}.
Hao Liu is supported by the National Natural Science Foundation of China (Grants No.~11653002, 11653003), the Strategic Priority Research Program of the CAS (Grant No.~XDB23020000) and the Youth Innovation Promotion Association, CAS.

\bibliography{newbib}

\end{document}